\documentstyle [aps,prb,twocolumn,epsfig]{revtex}
\begin{document}
\twocolumn[\hsize\textwidth\columnwidth\hsize\csname@twocolumnfalse%
\endcsname
%\draft

\title{Localization length in a random magnetic field}

\author{J. A. Verg\'es}
\address{Instituto de Ciencia de Materiales de Madrid,
Consejo Superior de Investigaciones Cient\'{\i}ficas,\\
Cantoblanco, E-28049 Madrid, Spain}

\date{\today}

\maketitle

\begin{abstract}
Kubo formula is used to get the d.c conductance of a statistical
ensemble of two dimensional clusters of the square lattice 
in the presence of random magnetic fluxes.
Fluxes traversing lattice plaquettes are distributed
uniformly between $- {1 \over 2} \Phi_0$ and ${1 \over 2} \Phi_0$
with $\Phi_0$ the flux quantum.
The localization length is obtained from the exponential decay
of the averaged conductance as a function of the cluster side.
Standard results are recovered when this numerical approach
is applied to Anderson model of diagonal disorder.
The localization length of the complex non-diagonal model of disorder
remains well below $\xi = {10}^4$ in the main part of the band
in spite of its exponential increase near the band edges.
\end{abstract}

]

\section{Introduction}

The two-dimensional (2D) motion of a single particle in
a random magnetic field is
attracting strong interest since the model is related both to
the slave-boson description of high $T_c$ superconductors\cite{slave}
and the Chern-Simons theory of the half-filled Landau level.\cite{chern}
The existence of recent experiments measuring 
transport properties in a static random magnetic
field adds considerable interest to this subject.\cite{exp}
In principle, all states of a 2D system are exponentially localized
according to the scaling theory of localization.\cite{localizacion}
The presence of a magnetic field that destroys
time-reversal symmetry makes a stand against localization.
Nevertheless,
perturbative renormalization group calculations show that all
states remain localized.\cite{sigma,zhang}

While theory points towards localization, numerical work
has not led to a consensus of opinion.\cite{numerics}
Loosely speaking, states {\it look} extended in
finite samples as soon as the edge of the band is left
and a definite answer is quite difficult.
Therefore, the most extended conclusion is that a mobility edge
separates localized states from extended or critical states.

The model of 2D random fluxes has been previously studied from a statistical
point of view.
The analysis of several statistical magnitudes (nearest-neighbor spacing
distribution, the probability distribution of wavefunction weights, etc.)
shows a systematic flow from the
Gaussian Unitary Ensemble (GUE) statistics at small system sizes
to Poisson statistics at larger sizes.\cite{verges}
If GUE and Poisson statistics are the only possibilities
in the macroscopic limit, localization over the whole spectrum follows
from the referred numerical results. If, on the contrary, more fixed points are
possible, then the flow could eventually start near the metallic universality
class and stop at certain {\it new} class characterized probably by
some fractal properties (for example, some power law increase of the
number of sites visited by an eigenstate).

Both the theoretical approaches to the metal-insulator transition and
the numerical tools used for the study of localization have a long history.
\cite{localizacion,efetov,lee,kramer}
The standard numerical method has been really
successful describing the critical behavior of the 3D metal-insulator
transition of Anderson model.\cite{kramer}
Nevertheless, the numerical procedure is somewhat artificial.
Let me remind that firstly,
localization lengths of 1D systems (strips or bars) of increasing
transverse size are calculated and, secondly, extrapolation of the data
gives the localization length of infinite 2D or 3D systems.
This introduces a fictitious symmetry breaking of
the 2D or 3D system since a particular direction is preferentially treated.
Moreover, being all 1D states exponentially localized, they become
eventually extended after doing finite-size scaling analysis of the data.
This last part of the calculation introduces some uncertainties in the
procedure.

In this work, I return to the original idea of {\it measuring} the conductance
of finite samples on the computer. \cite{inicios1,inicios2}
Today's use of the older scheme has some advantages:
mesoscopic physics has led to a more efficient use of
Kubo formula\cite{datta},
computers are much more powerful,
there is a large experience in the numerical calculation of localization
lengths, etc..
Based on the statistical analysis of conductance data obtained
for samples of increasing size, the {\it direct} evaluation of the localization
length is possible. I find this approach really appealing for two reasons:
i) the {\it exact} static conductance is obtained via Kubo formula (neither
a small imaginary part should be added to the energy nor other
numerical trick should be used to reach the static limit of the conductance
of a finite sample), and
ii) the scaling behavior of a {\it measurable} magnitude (conductance)
is directly analyzed.

The paper is organized as follows. The model that is
numerically studied in the paper is defined in Section II,
some remarks about the use of the Kubo formula
within the present context are given in Section III,
the size dependence of the conductance is discussed in Section IV,
results for the Anderson model and the model with random fluxes are
presented in Section V, and finally, Section VI ends the paper with some
remarks.

\section{Lattice model}

The Hamiltonian describing random magnetic fluxes on a $L \times L$ cluster
of the square lattice is :

\begin{equation}
\hat H = \sum_{l=1}^{L^2} \epsilon_l \hat c_{l}^{\dag} \hat c_{l}
         - ~t {\sum_{<l l'>}}
        {e^{2 \pi i \phi_{l l'}}
        \hat c_{l}^{\dag} \hat c_{l'}}
        ~~,
\label{RMF}
\end{equation}

\noindent
where $\hat c_{l}^{\dag}$ creates an electron on site $l$,
$l$ and $l'$ are nearest-neighbor sites, 
$\epsilon_l$ are random diagonal energies chosen with
equal probability from the interval $[-W/2,W/2]$,
and $- t$ is the hopping energy (hereafter $t=1$ is chosen).
The flux through a given loop $S$ on the lattice is
$\Phi_S = \sum_{<l,l'> \in S} \phi_{l l'}$
measured in units of flux quantum $\Phi_0 = {{hc} \over e}$.
Link fields satisfy $\phi_{l l'}=-\phi_{l' l}$.
Either link fields (Meissner phase) or magnetic fluxes (Debye phase)
can be randomly selected.\cite{wheatley}
Here, the second possibility has been chosen:
uncorrelated fluxes are randomly selected with equal probability
from the interval $[- {1 \over 2},{1 \over 2}]$ while $W=0$.
Notice that we are considering the model with pure non-diagonal disorder
that shows maximum magnetic disorder.
This model will be called
RMF (Random Magnetic Fluxes or Random Magnetic Field) model hereafter.

Some additional calculations have been done
for a {\it real} version of Hamiltonian (\ref{RMF}).
As before, diagonal disorder is absent ($W=0$) but now link
field values are restricted to $0$ and $1/2$ that are randomly chosen with
equal probability. The model describes
a square lattice in which the hopping integral is constant
in absolute value but takes random signs.
This model will be shortly called RHS (Random Hopping Signs) model.
Finally, a pure Anderson model of real diagonal disorder has also been studied
in order to compare with results in the literature.
In this case, $W \neq 0$ and all link fields vanish.

\section{D.c. conductance via Kubo formula}
According to Kubo's linear response theory,
the static electrical conductivity is given by:\cite{nozieres}
\begin{equation}
{\mathrm G} \equiv \sigma_{xx}(0) = - ({{e^2} \over h})
2 {\mathrm Tr} [ (\hbar \hat v_x) {\mathrm Im} {\mathcal G}(E)
(\hbar \hat v_x) {\mathrm Im} {\mathcal G}(E) ] ~~~,
\label{Kubo}
\end{equation}
where the velocity (current) operator $\hat v_x$ is related to the position
$\hat x$ operator by the equation of motion:
\begin{equation}
i \hbar \hat v_x = [\hat H , \hat x]
\label{vx}
\end{equation}
and ${\mathrm Im} {\mathcal G}(E)$
is defined in terms of advanced and retarded Green functions as:
\begin{equation}
{\mathrm Im} {\mathcal G}(E) \equiv {1 \over {2i}}
[{\mathcal G}^+(E) - {\mathcal G}^-(E)] ~~~.
\label{img}
\end{equation}

The conductance is calculated for a system consisting of a
$L \times L$ cluster of the square lattice connected to two ideal
semiinfinite leads of width $L$.
The exact form of the electrical field does not matter
in a linear response one-electron theory. Therefore,
an abrupt potential drop at one of the two cluster sides provides the simplest
numerical use of the Kubo formula. In this case, operator $\hat v_x$
has finite matrix elements on only two adjacent layers (see Eq.(\ref{RMF}))
and Green functions are just needed for this considerably restricted
subset of sites owing to the trace appearing in Eq.(\ref{Kubo}).
Since Green functions are given by:
\begin{equation}
[E - \hat H - \hat \Sigma_{\mathrm r}(E) - \hat \Sigma_{\mathrm l}(E)]
{\mathcal G}(E) = I ~~~,
\label{green1}
\end{equation}
where $\hat \Sigma_{\mathrm r}(E)$ and $\hat \Sigma_{\mathrm l}(E)$
are the selfenergies introduced
by the semiinfinite right and left leads, respectively\cite{datta},
the evaluation of the desired elements of ${\mathrm Im} {\mathcal G}(E)$
is efficiently achieved applying an $LU$ decomposition to the band matrix
\begin{equation}
<l|E - \hat H - \hat \Sigma_{\mathrm r}(E) - \hat \Sigma_{\mathrm l}(E)|l'> ~~~.
\label{green2}
\end{equation}

Selfenergies are troubleless calculated for any energy $E$
using a recurrent algorithm.
Retarded or advanced Green functions
(${\mathcal G}^+(E)$ or ${\mathcal G}^-(E)$)
are obtained using the corresponding selfenergies in Eq.(\ref{green1}).
Since lead selfenergies add an imaginary part to some diagonal elements
of the Hamiltonian,
there is no need to add a small imaginary part to
the energy $E$ before solving the set of linear algebraic equations.

The discussion of a technical point is here in order. It has been proved
that Kubo formalism leads to the more familiar Landauer-B\"uttiker
expression of the conductance as a sum of transmission coefficients over
channels. \cite{equivalencia}
Usually, transmission is obtained within a transfer matrix
technique in which the selfenergy produced by the lead at the left is
iterated through the disordered
part of the sample and matched at the right side to the selfenergy coming
from the right lead. In this calculation, matrices of dimension $L$ are
inverted $L$ times. It could seem that this method is computationally
more appropriate
because it deals with smaller matrices (the elements of the Green function
within the sample are given by the inverse of matrix (\ref{green2})
which has dimension $L^2$). Nevertheless, since matrix (\ref{green2})
is actually a band matrix of bandwidth $2L+1$ and only a few elements of the
Green function are necessary for the evaluation of the conductance via
Kubo formula, standard numerical procedures ($LU$ decomposition of a band
matrix followed by both a forward substitution and a backsubstitution for
each needed Green function element) are quite advantageous. Moreover,
canned subroutine packages customized for particular computers can be used.
Thus, we end with a very simple computer code written in a high level
language that can be advantageously run on any given computer.
Typically, 127 seconds are needed for the computation
of the conductance of a $100 \times 100$ sample on an alpha DECstation 3000.

\section{Scaling of the conductance}
Conductance is {\it measured} for a statistical ensemble of $L \times L$
clusters with random disorder. This provides a set $\{ {\mathrm G}_i \}$
of $N$ values of the static conductance that allows the evaluation of both
the average value of conductance
\begin{equation}
{\overline {\mathrm G}} = {1 \over N} \sum_{i=1}^N {\mathrm G}_i
\label{mean}
\end{equation}
and its variance
\begin{equation}
{\mathrm Var} ({\mathrm G}_1 ...{\mathrm G}_N) = {1 \over N-1}
\sum_{i=1}^N ({\mathrm G}_i - \overline {\mathrm G})^2.
\label{variance}
\end{equation}
The error bar of $\overline {\mathrm G}$ is given by its standard deviation
\begin{equation}
\sqrt{<{\overline {\mathrm G}}^2> - {<\overline {\mathrm G}>}^2}
\label{errorbar}
\end{equation}
which is well approximated by
\begin{equation}
\sigma(\overline {\mathrm G})=
\sqrt {{{\mathrm Var} ({\mathrm G}_1 ...{\mathrm G}_N)} \over N}
\label{errorbarbis}
\end{equation}
as soon as $N$ is large enough.
Since conductance shows typical universal fluctuations of order 1 (in
units of $e^2/h$), the error of the measured average conductance is
$\sim N^{-{1 \over 2}}$.

This way of characterizing the whole distribution by just two values
($\overline {\mathrm G}$ and its error bar) works fine for magnitudes
showing an almost Gaussian statistics. Nevertheless, the distribution of
conductances clearly deviates from normal when the regime of exponential
localization is being approached. In this regime both theory and numerics
suggest that the distribution of the logarithm of the conductance is the
appropriate statistical variable showing a normal distribution around its
average value. Therefore, the usual practice is the analysis of
$\log {\mathrm G}$ whenever the distribution of $\mathrm G$ deviates
significantly from normal. I am not going to follow this standard way
of analyzing the conductance for several reasons:
i) In principle, since $\mathrm G$ is measured, it seems preferable to
work directly with it. Moreover, $\mathrm G$ is both bounded from below
($0$) and from above ($L$, the number of channels). While conductance will
vanish in the localization regime, its logarithm will diverge.
Consequently, the variance of $\mathrm G$ is bounded but the variance of
$\log {\mathrm G}$ is not.
Computationally, this means that logarithm averages show more {\it noise}
in finite sampling.
ii) The statistical characterization of the conductance should work from
the metallic regime to the localized one since regime can change as a
function of the system size. The switch from $\mathrm G$ to
$\log({\mathrm G})$ is somewhat ambiguous.
iii) The median plays quite good the role of a central value of some
statistical magnitude even when the distribution of this magnitude is
not Gaussian. For example, I have checked that the conductance median
$\mathrm G_{med}$ gives exactly the same exponentially decaying behavior as
$\overline {\log({\mathrm G})}$ in the localization regime. Actually, the
median will work for any variable change that preserves ordering among data,
i.e., if $f({\mathrm G}_i) < f({\mathrm G}_j)$ whenever
${\mathrm G}_i < {\mathrm G}_j$.
This property is quite general and makes the use of the median quite robust.
iv) Unfortunately, the median has two statistical disadvantages. First, even
for a normal distribution of the original variable, its variance for a finite
sampling is larger that the corresponding variance of the mean
value.\cite{cramer} Second, the estimation of its error bar is not as
straightforward as the calculation of the standard deviation of
the mean (Eq.\ref{errorbarbis}).
Actually, one is forced to divide the database into subsets, obtain several
median estimates and calculate their variance. The result is a bad estimation
of the uncertainty of the median.
v) In conclusion, I will use the conductance median {\it without} error
bar as the magnitude giving the central value of the
conductance distribution whenever a transition from
the metallic regime to the localization regime takes place somewhere within
the computation. On the other hand,
the use of the mean value and its error bar will be preferred
when the conductance distribution remains normal.
From here on, either $\overline {\mathrm G}$ or $\mathrm G_{med}$ 
will be referred as conductance $\mathrm G$
and will be given in units of the conductance quantum $e^2/h$.

Localization is an asymptotic property of the conductance as a function of
the linear cluster size $L$. In order to correctly describe this behavior
using measurements done on finite samples, finite size corrections
should be allowed. Notice that the conductance of a 2D square cluster of
side $L$ increases linearly in the absence of disorder; in fact, the
conductance is an integer value that gives the total number of channels
that are open at the Fermi energy
(this number is exactly $L$ at the band center).
The following scaling models have been used in this work:
\begin{equation}
{\mathrm G} (L) = (a+bL^{-1}+cL^{-2}+ ...)
\exp (-{{2 L} \over \xi}) ~~~~,\\
\label{fit}
\end{equation}
and
\begin{equation}
{\mathrm G} (L) = {L \over {a+bL}}
\exp (-{{2 L} \over \xi})   .
\label{fit_coc}
\end{equation}
In both cases, the term multiplying the exponentially decaying factor
is designed to describe finite size effects.
$\xi$ measures the localization length.
The first expression is more
intuitive and works always fine (with a large enough number of parameters)
while the second expression describes better the metallic regime in which
the conductance increases linearly with the system size $L$.
All the results presented in the next Section allow a good fit with just
three parameters ($a, b, \xi$). The localization length does not depend
sensibly on the used model.

Finally, let me comment about the reliability of the localization
length obtained by the finite scale modeling. Generally speaking,
parameters $a, b, c, ...$ are determined by conductance values
of small cluster sizes while the localization length $\xi$ is determined by
the behavior of conductance at larger system sizes. Since conductance starts
decreasing after a region of initial increase,
the estimation of the localization lengths is reliable only if 
the crossover to the region in which conductance decreases can be reached
by numerical simulation. According to the data presented in this paper,
this happens for localization lengths that are well below ${10}^4$ (lengths
are everywhere given in units of $a$, the lattice constant).
Notice that the existence of localization at large $L$ is built in the
model (see Eqs.(\ref{fit}--\ref{fit_coc})): a decaying conductance does not
necessarily mean an exponentially decaying law.
In other words, numerical simulation does not prove localization but
gives a measure of its characteristic length $\xi$.
Nevertheless, after trying a lot
of alternative (and worser, i.e., with larger $\chi^2$ deviations from data)
models for the description of the conductance dependence
on system size, one gets more or less convinced that exponential decay at
large $L$ values is quite plausible.\cite{elihu}

\section{Results}

Although my primary interest is the determination of localization
lengths for Hamiltonians describing random magnetic fluxes, the paradigmatic
Anderson model of diagonal disorder has also been studied in order to
check the numerical tools used in this work.
Conductance values are everywhere given in units of $e^2/h$.

\subsection{Anderson model}

Hamiltonian (\ref{RMF}) without link fields ($\phi_{l l'} = 0$ for all $l, l'$
pairs) describes the Anderson model of diagonal disorder. The width $W$
of the diagonal energies distribution measures the degree of disorder.
Exponential localization of all wavefunctions is the accepted state of the
art for this model. Nevertheless, measuring the localization length when
disorder is weak ($W < 2$) is numerically very difficult.
Wavefunctions are clearly localized in the band tails but {\it look} extended
well within the band. This fact has originated more than one claim pointing
to the existence of a metal-insulator transition at some critical energy
(mobility edge) within the band.

Following tradition, I have collected conductance data at the band center
($E=0$). It is assumed that states at finite energies are more localized
than the band center one. Fig. 1 shows the behavior of both the mean
and the median of the conductances samples as a function of the side $L$
of the $L \times L$ disordered cluster in which the current is measured.
Since disorder is very large ($W=10$), both measures of a central value show
a clear exponential decay. Nevertheless, there are also important differences:
localization lengths do not coincide and mean values show more dispersion,
i.e., they do not follow a straight line as
good as the median does. Looking closer to the data, the reason for the
disagreement is obvious: conductance is very small ($G \ll 1$), its 
distribution is not normal and, consequently, the mean is dominated by outliers
conductances of a relatively large value. On the other hand, the logarithm
of the conductance seems to behave statistically good. As said in Section IV,
the median of $\mathrm G$
gives a good estimation of the central value of its statistical distribution
if any monotonic function of $\mathrm G$ behaves normally.
Although the statistical distribution of the conductance becomes almost
Gaussian for smaller degrees of disorder,
I have systematically used the median in the
rest of this Section in order to be completely self-consistent.

Figures 2 and 3 show the evolution of the scaling behavior of the conductance
as disorder diminishes. While finite size effects are still unimportant
for $W=5$, a region in which conductance increases linearly with the
system size has clearly developed for $W=2$.
This region extends more and more as $W$ approaches zero.
Actually, the inference of a localization length
beyond $W=2$ is not possible analyzing conductance data for sample sizes
below $L=100$. As commented before, only systems having a localization
length below $\xi \sim {10}^4$ show numerical indications of
localization for sample sizes of the order of $L = 100$. In other words, when
$\xi \gg L$ one just obtains the linear increase of conductance due to the
linear increase of transport channels.

Fig. 4 summarizes the main results of this subsection: localization length is
plotted as a function of disorder. It seems clear that $\xi$ diverges for
vanishing disorder. In fact, a $\chi^2$ fit by power law gives an exponent
very close to $-4$ and even the simpler fit by $1+a W^{-4}$ works pretty good
(see Fig. 4). Localization length values are also compared with results
obtained by finite size scaling based on data collected for strips of
increasing width.\cite{anderson1,anderson2,anderson3}
Surprisingly, both methods do not give identical results. Present values of
$\xi$ are somewhat larger for large disorder but their evolution for smaller
disorder values is not so steep. For small disorder values,
the scaling of the conductance of square clusters attached to
semiinfinite wires yields considerably smaller localization lengths.
I have no explanation for this discrepancy.
It could be that numerical extrapolation of $\xi$ from finite
strips has a larger uncertainty or even that the whole procedure is dominated
by the more localized states. Let me repeat here that $\mathrm G$ is
a statistical variable and that the scaling behavior of a {\it single}
magnitude used to characterize the whole distribution of values
can certainly depend on the choice (see Fig. 1 for an example of an extreme
discrepancy).

\subsection{Random Magnetic Flux}

This part presents the main results of the paper. Let me begin with a
detailed comparison between statistics based on the mean of the conductance
distribution and statistics based on its median. Fig. 5 compares in a
logarithmic scale the behavior
of both magnitudes at an energy $E=-3.3$ close to the band limit.
The values shown for $L=100$ are not so precise as the rest because
a sensible percentage of the samples have conductances that are smaller
than numerical precision.
Apart from minor size effects at small cluster
sizes, both magnitudes show exponentially decaying behavior.
While localization
does not offer doubts, the exact value of localization length depends on
the magnitude used to characterize the central value of the distribution of
conductances. $\xi$ increases from $20.9$ for the median to $60.6$ for the
mean. As said previously, the first value is the correct one since the
distribution of the conductance logarithm is indeed normal and gives the same
median. Fig. 6 shows that the problem becomes softer as soon as energies
well within the band are studied. Although conductance is still below one
quantum unit, its distribution comes closer to a Gaussian distribution and
therefore, mean and median do not differ as much as in the previous example.

Let us analyze now the trend followed by the size dependence of conductance
when energy is varied.
While localization is clear near the band edges, it is quite difficult
to get some insight into the scaling behavior of conductance at energies
$|E| < 3$. Fig. 7 shows my larger data collection corresponding to
an arbitrary energy $E=0.1751$ close to the band center. 5500
{\it measurements} of the conductance have been necessary to reduce the
error bar of the mean to the small values shown in the Figure for $L < 70$.
At larger sizes, a still very large number of $500$ conductances
of randomly generated samples have been calculated. Error
bars are considerably larger but not so much to preclude a precise analysis.
In my opinion, the decrease of the conductance
for increasing system sizes does not offer any doubt.
Moreover, a special calculation at $L=200$ has been carried out
using a different algorithm that allows the study of larger systems paying
the price of longer computer times.\cite{gauss} The decrease of the conductance
below its value at $L \sim 100$ is quite clear:
$\mathrm G(200)=1.330 \pm 0.014 < \mathrm G(100)=1.381 \pm  0.005$.

The good quality of the
database for this energy gives the opportunity of doing a reliable fit
by some chosen law. Using Eq.(\ref{fit}) with four parameters ($a,b,c,\xi$),
a localization length of $\xi = 5152.8$ is obtained.
Fig. 7 shows the fitted model as a thick
continuous gray curve. The relatively large value of the
localization length is just below the threshold of {\it visibility} ---given
the present computer facilities.
Let me insist once again, that the data shown in Fig. 7 do not prove
exponential decay of conductance although they are nicely described by
an exponential law including finite size corrections.
On the other hand,
it is obvious that any law with properly selected parameters is able
to reproduce the roughly linear decrease of conductance for $L>50$.
Nevertheless, reproducing the whole scaling behavior is not so trivial.
For example, if all conductance values except the one corresponding
to $L=200$ are fitted by Eq.(\ref{fit}) and by a rational function
model like:
\begin{equation}
{\mathrm G} (L) = {{aL+bL^2} \over {1+cL+dL^2+eL^3}} ~~~,
\label{fit_fraccion}
\end{equation}
the fit fails to reproduce the last point if the rational function is used
---both adjusting ($a,c,d$) (thin continuous line of Fig. 7) and
adjusting ($a,b,c,d,e$) (thick dashed line of Fig. 7)---
whereas it remains almost the same when the exponentially decaying model
is employed (thick continuous gray line of the same figure).

Actually, a large database has been also collected for the conductance
distribution at the band center. Mean conductance steadily increases
up to the largest calculated size ($L=200$). A very large localization length
$\xi = 16707.5$ is obtained using Eq.(\ref{fit}) with four parameters,
although numerical uncertainty is as large as its value.
The steady increase of the conductance at the band center of this model
was firstly obtained in the seminal work of Lee and Fisher. \cite{inicios2}
Although they did, in principle, the same kind of computation,
numerical values do not coincide for some unknown reason (the
inset of Fig. 2 of the referred paper shows conductances that are about
seven times smaller than the ones obtained in this work).
In any case,
either $\xi$ is infinite just at the band center or larger than
any other calculated localization length.

Let me point, that the mean conductance has been systematically used to get
the localization length of this model for several energies within the band.
Numerically, the advantage of the mean relative to the median is its
better normal distribution for finite samples.\cite{cramer}
Its use is justified as soon as conductance distribution behaves normally
(Gaussian distribution).  Fig. 8 shows that this is indeed the case
at $E=0.1751$ both for $L=64$ and for $L=200$.
The small decrease of the mean conductance is not distinguishable in the
distribution. This figure also shows the typical
width of about one quantum unit that has been observed for the conductance
fluctuations in numerous occasions (the standard deviation of the
distribution obtained for $L=200$ equals $0.314$).

Fig. 9 collects the localization length obtained at several energies. The
almost exponential increase of $\xi$ near the band edges perfectly reproduces
the findings of previous studies. \cite{numerics} On the other hand, the
localization length seems to reach a saturation value below $10^4$ in the
larger part of the band being the band center the only possible exception.
Two comments are opportune at this point.
First, a continuous logarithmic increase of the localization length up to
the band center implies a more than macroscopic value of $\xi$ for almost
all states in the band. If localization is much larger than the sample size,
the own concept of localization looses its meaning. In my opinion, the scenario
shown in Fig. 9 is much more attractive: localization length is very large
but not {\it infinite}, that is, irrelevant.
Second, although renormalization group theories also predict
$\xi \sim l \exp (A l^2)$
where $l$ is the mean free path and $A$ some finite constant,
my numerical results do not necessarily contradict this behavior because
the RMF model gives values of $l$ of order of $1$
when non-diagonal disorder is maximum. \cite{mfp}

\subsection{Random Hopping Sign}

The last case studied in the paper corresponds to a model of real
hopping with a random sign. The Hamiltonian is given by (\ref{RMF}) when
link fields are limited to values $0$ and $1 \over 2$. 
The band extends from $\sim -2 \sqrt 3$ to $\sim 2 \sqrt 3$ in this case.
Being the amount of
non-diagonal disorder fixed, conductance can be studied as a function of
energy. Since this model belongs to a universality class that is different
from the previous one (orthogonal instead of unitary), some differences
in the scaling behavior of conductance could, in principle, be expected.
Nevertheless, numerical results show only minor changes.

Scaling of conductance is shown
in Fig. 10 for two different energies: $E=0$ (band center)
and an arbitrary energy $E=0.1751$ close to it.
While localization length is almost infinite
($\xi = 23317$) in the first case, it is not very large in the second one
($\xi = 461.7$).
As in the complex model, an infinite value of $\xi$ is compatible
with conductance data at the band center but not with the clear decrease
shown by the conductance at the second energy.
In both cases, the mean and its error bar have been used because the
distribution of conductance values follows a Gaussian law in spite of its small
average value (Mean conductance is of the order of $1$ quantum unit
when the conductance of an ordered sample is as large as
$100$ at the band center of a cluster of side $L=100$).
Conductance scaling  at other energies follows a
trend that is quite similar to the one followed by the previously analyzed
RMF model. It can be concluded that the
universality class of the model does not matter noticeable in the scaling
behavior of conductance although numerical values of the localization
length are quite different ($\xi=461.7$ ($5152.8$) for RHS (RMF) model at
$E=0.1751$).

\section{Discussion}

Conductance sampling as a function of the system size has been methodically
used to infer the localization lengths of a variety of disorder models.
Results for the Anderson model of diagonal disorder are globally consistent
with the ones obtained using well known approaches.
Nevertheless, numerical values of $\xi$ are
somewhat surprising: although localization length diverges for vanishing
amount of disorder, the divergence is weaker than the one provided
by other methods. The origin of the discrepancy is not known although it
may be due to the non-normal distribution of conductances when the
localization regime is approached.

Both the real ---belonging to the Gaussian Orthogonal Ensemble universality
class--- and the complex (GUE universality class)
models of non-diagonal
disorder studied in this work show decaying values of the central
conductance as a function of cluster side at all studied energies.
When fitted with an exponential law corrected for finite-size effects,
values of the localization lengths are obtained. Although large they are
typically below $10^4$, a value that is the maximum localization length
giving a conductance that diminishes within the size values presently
accessible. The behavior of $\xi$ as a function of energy is remarkable.
After a region near the band edges in which $\xi$ increases exponentially,
a region of almost constant localization length follows.
Nevertheless, isolated energies giving diverging values of $\xi$
cannot be discarded by this procedure. Actually, the band center is a candidate
for showing a diverging localization length.

In my opinion, this work should motivate experimental research in the
mesoscopic transport field directed to the determination of the
localization length of disordered samples. As shown by the large amount
of data presented in this work, the decay of the conductance starts
at system sizes well below the localization length. This fact makes
the experimental observation of localization relatively easy. Only
after having experimentally related values of disorder (or what is
equivalent, values of the mean free path) with values of the localization
length, could localization theory make a major step from conjecture
to fact. On the other hand, the correct microscopic description
of the conductance scaling at a finite temperature would greatly expedite
a detailed comparison between experiment and numerics.

\acknowledgments
I acknowledge Luis Mart\'{\i}n-Moreno who helped me checking the computer
code used in this work running his own program based on the transfer
matrix technique for the same disordered system.
I also acknowledge interesting conversations with Luis Brey.
This work has been partially supported by Spanish 
Comisi\'on Interministerial de Ciencia y Tecnolog\'{\i}a (grant MAT94-0058-C02).

\newpage
\begin{figure}
\epsfig{figure=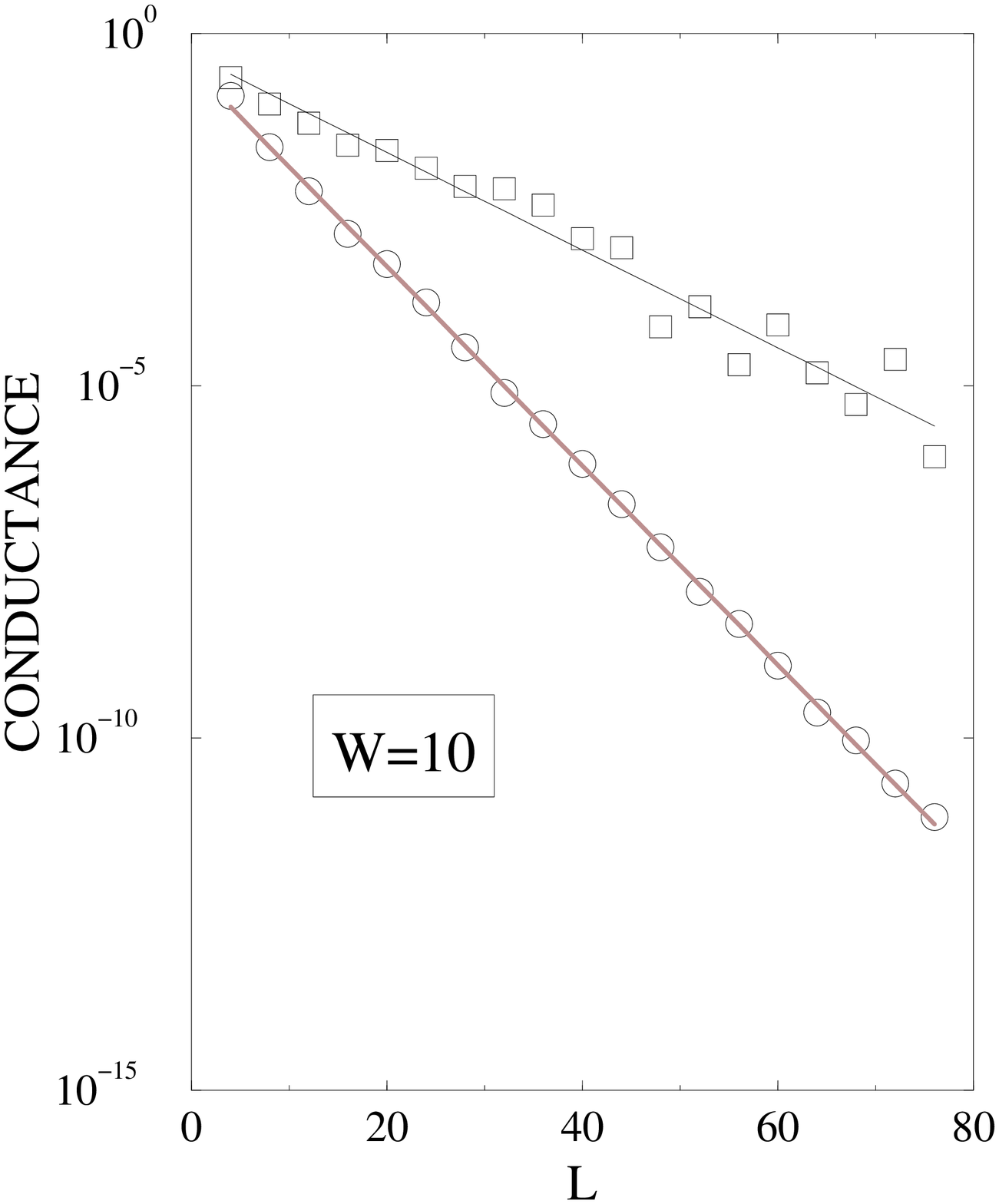,height=10cm,width=8cm}
\caption{Scaling of conductance for the Anderson model of diagonal
disorder with $W=10$. The median of the conductance distribution is shown
by circles in a logarithmic scale while the means obtained from the same
samples are shown by squares. Simple linear regression provides
a localization length of $\xi=6.1$ for the median and $\xi=11.1$ for
the mean (continuous lines).}
\label{Figure1.ps}
\end{figure}

\newpage
\begin{figure}
\epsfig{figure=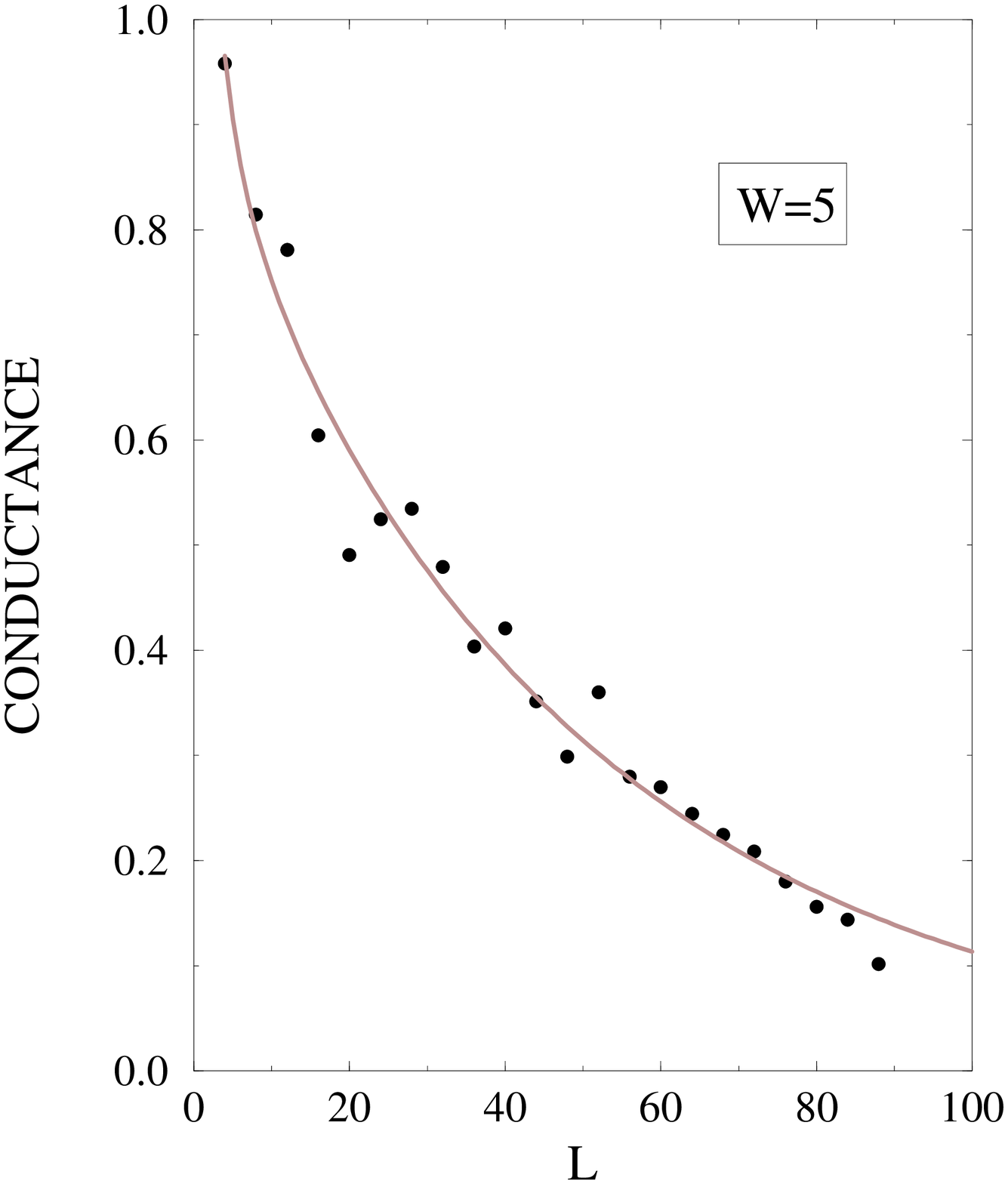,height=10cm,width=8cm}
\caption{Scaling of conductance for the Anderson model of diagonal
disorder with $W=5$. The median of the conductance distribution is shown
by circles. Fitting by Eq.(\ref{fit_coc})
gives a localization length of $\xi=98.8$ (continuous line).}
\label{Figure2.ps}
\end{figure}

\newpage
\begin{figure}
\epsfig{figure=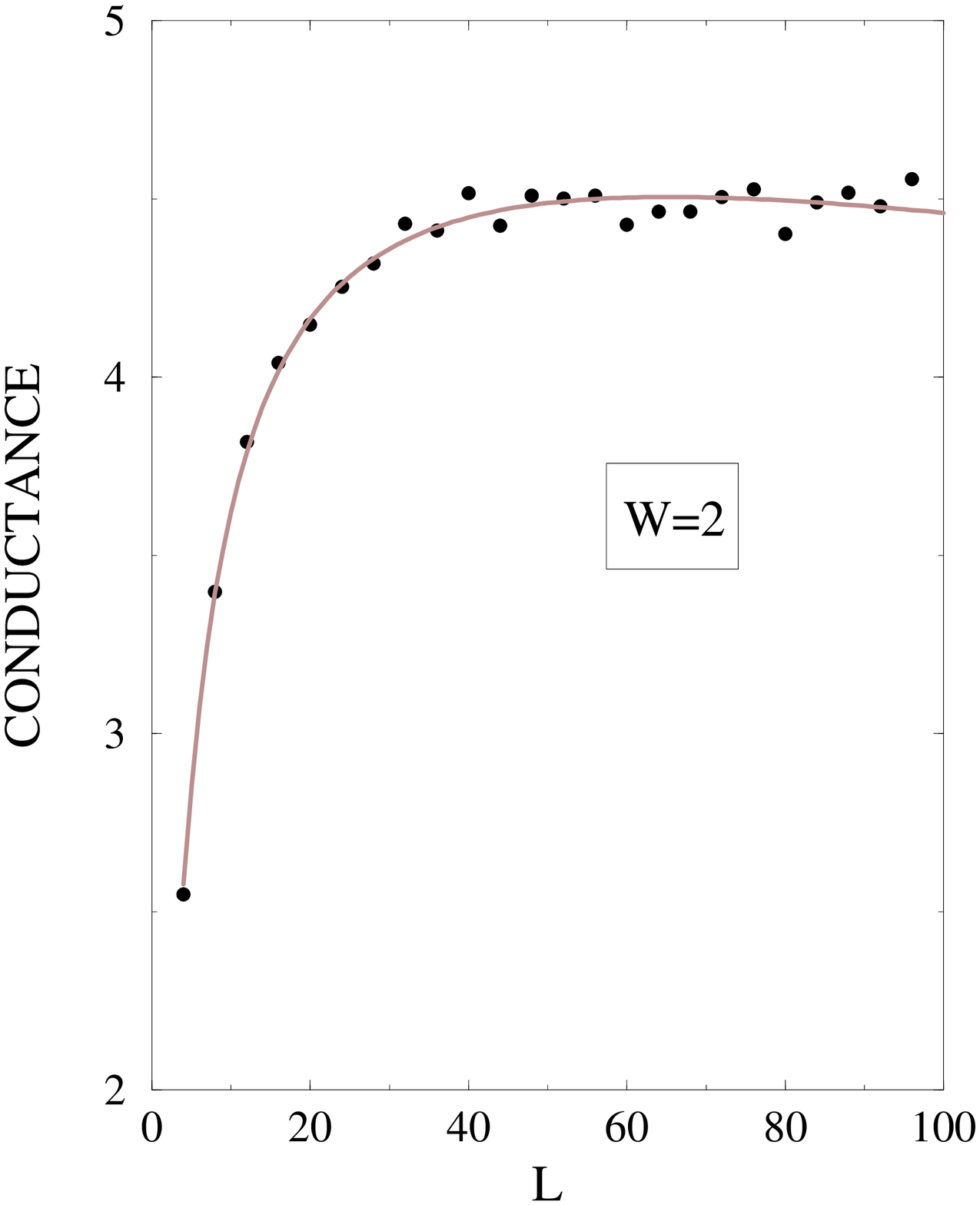,height=10cm,width=8cm}
\caption{Scaling of conductance for the Anderson model of diagonal
disorder with $W=2$. The median of the conductance distribution is shown
by circles. Fitting by Eq.(\ref{fit_coc})
gives a localization length of $\xi=2382.0$ (continuous line).}
\label{Figure3.ps}
\end{figure}

\newpage
\begin{figure}
\epsfig{figure=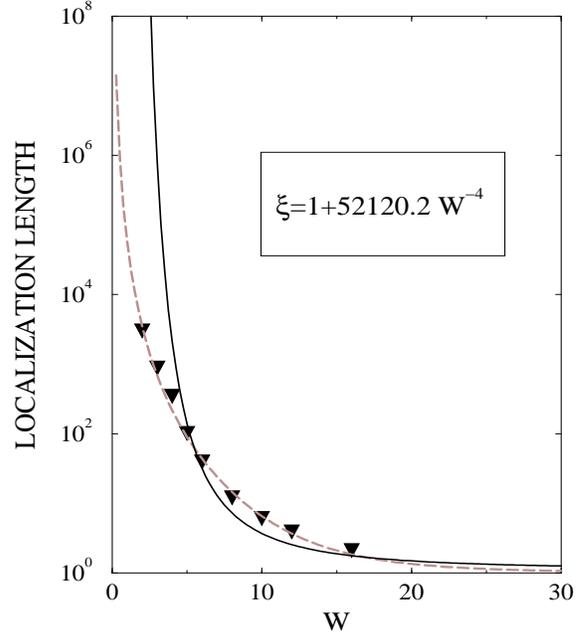,height=10cm,width=8cm}
\caption{Localization length for the Anderson model of
diagonal disorder. Triangles give my estimates, the dashed line shows
the approximate law written in the box, and the continuous line plots
$1.1 \exp [(11/W)^2]$, the behavior given in Ref.\cite{anderson3}.
Localization length is measured in units of the lattice constant and
the width $W$ of the distribution in units of the hopping energy $t$.}
\label{Figure4.ps}
\end{figure}

\newpage
\begin{figure}
\epsfig{figure=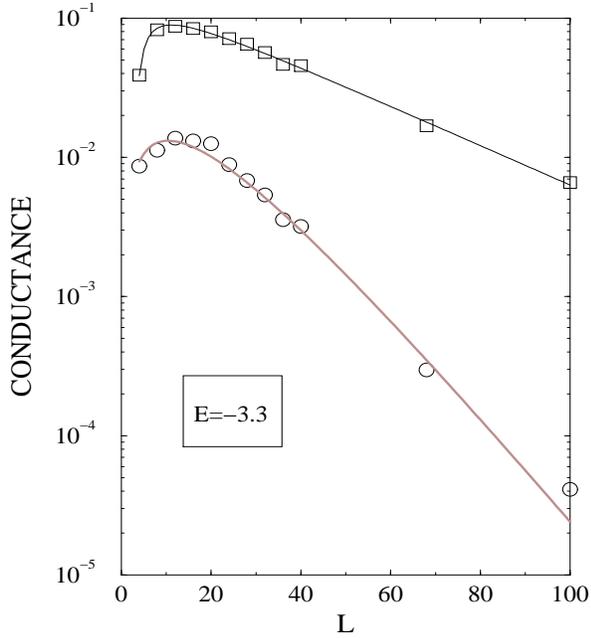,height=10cm,width=8cm}
\caption{Scaling of the conductance for the Random Magnetic
Flux model at an energy $E=-3.3$ close to the band edge.
Circles give the median of the conductance distribution evaluated for
samples including 5000 data while squares show the mean of the same
distributions. Localization length changes from $\xi=20.9$ for the
median to $\xi = 60.6$ for the mean.}
\label{Figure5.ps}
\end{figure}

\newpage
\begin{figure}
\epsfig{figure=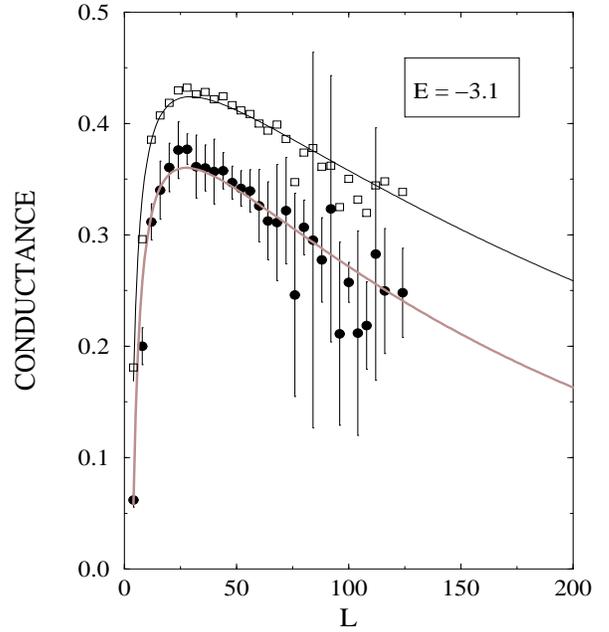,height=10cm,width=8cm}
\caption{Scaling of the conductance for the Random Magnetic
Flux model at an energy $E=-3.1$ not so close to the band edge.
Circles show median values with a rough error estimate while squares show
the evolution of the conductance mean. In the last case, the error bar is of
the order of the symbol.
The clear decay corresponds to a localization length of $\xi = 378.4$
for the median and $\xi = 592.5$ for the mean.}
\label{Figure6.ps}
\end{figure}

\newpage
\begin{figure}
\epsfig{figure=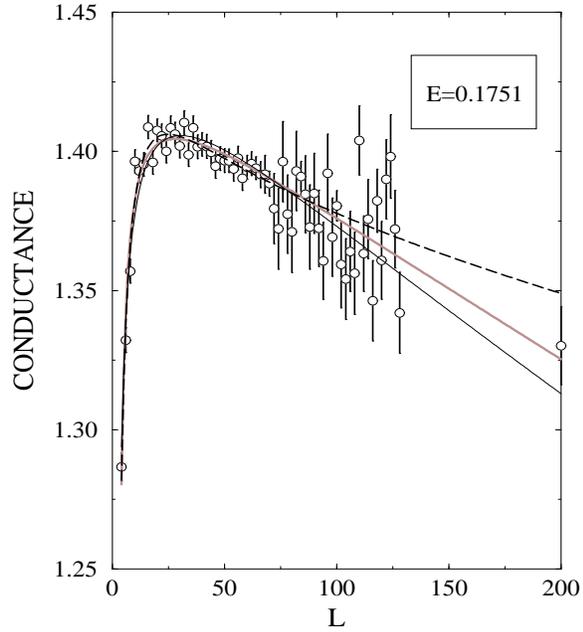,height=10cm,width=8cm}
\caption{Scaling of the conductance for the Random Magnetic
Flux model at energy $E=0.1751$ close to the band center. 5500
"measurements" of the conductance have been possible up to
$L=70$. From this size on, {\it only} 500 define the average value
with a larger error bar. The average conductance at $L=200$
has been calculated using a different numerical inversion subroutine.
Finite size scaling with Eq.(\ref{fit}) gives $\xi=5152.8$.}
\label{Figure7.ps}
\end{figure}

\newpage
\begin{figure}
\epsfig{figure=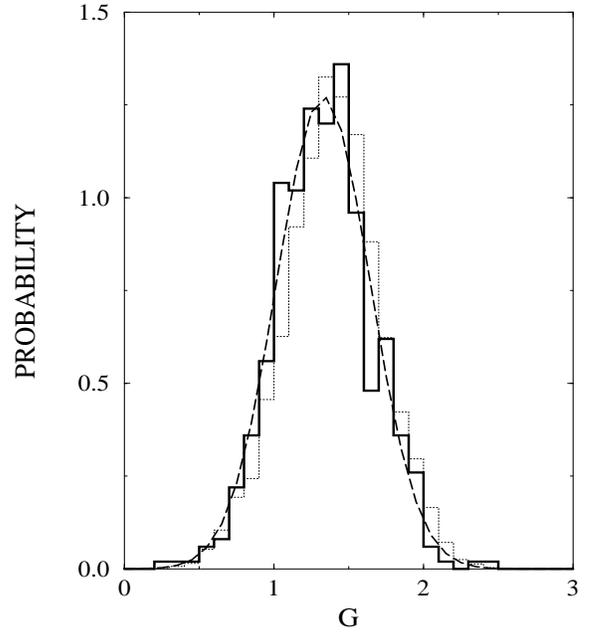,height=10cm,width=8cm}
\caption{Conductance distribution of the Random Magnetic Flux
model at energy $E=0.1751$ for two different sizes:
i) $L=64$ (dotted line) and ii) $L=200$ (thick continuous line).}
\label{Figure8.ps}
\end{figure}

\newpage
\begin{figure}
\epsfig{figure=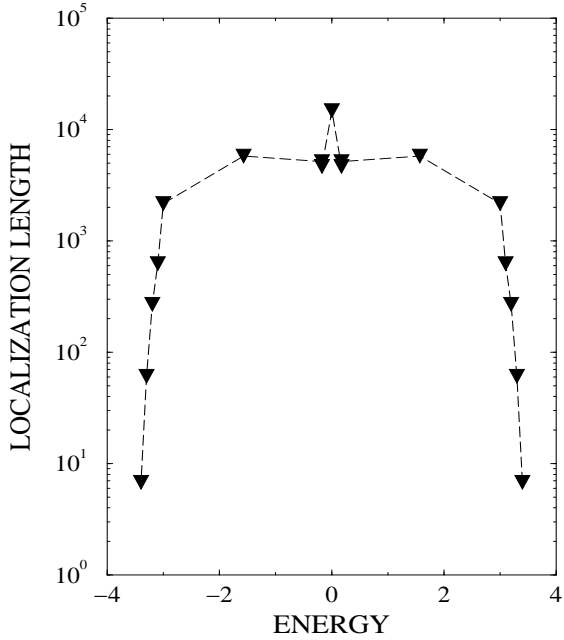,height=10cm,width=8cm}
\caption{Localization length for the Random Magnetic Flux
model as a function of energy. The band limits are approximately
at $\pm 2 \sqrt 3$.}
\label{Figure9.ps}
\end{figure}

\newpage
\begin{figure}
\epsfig{figure=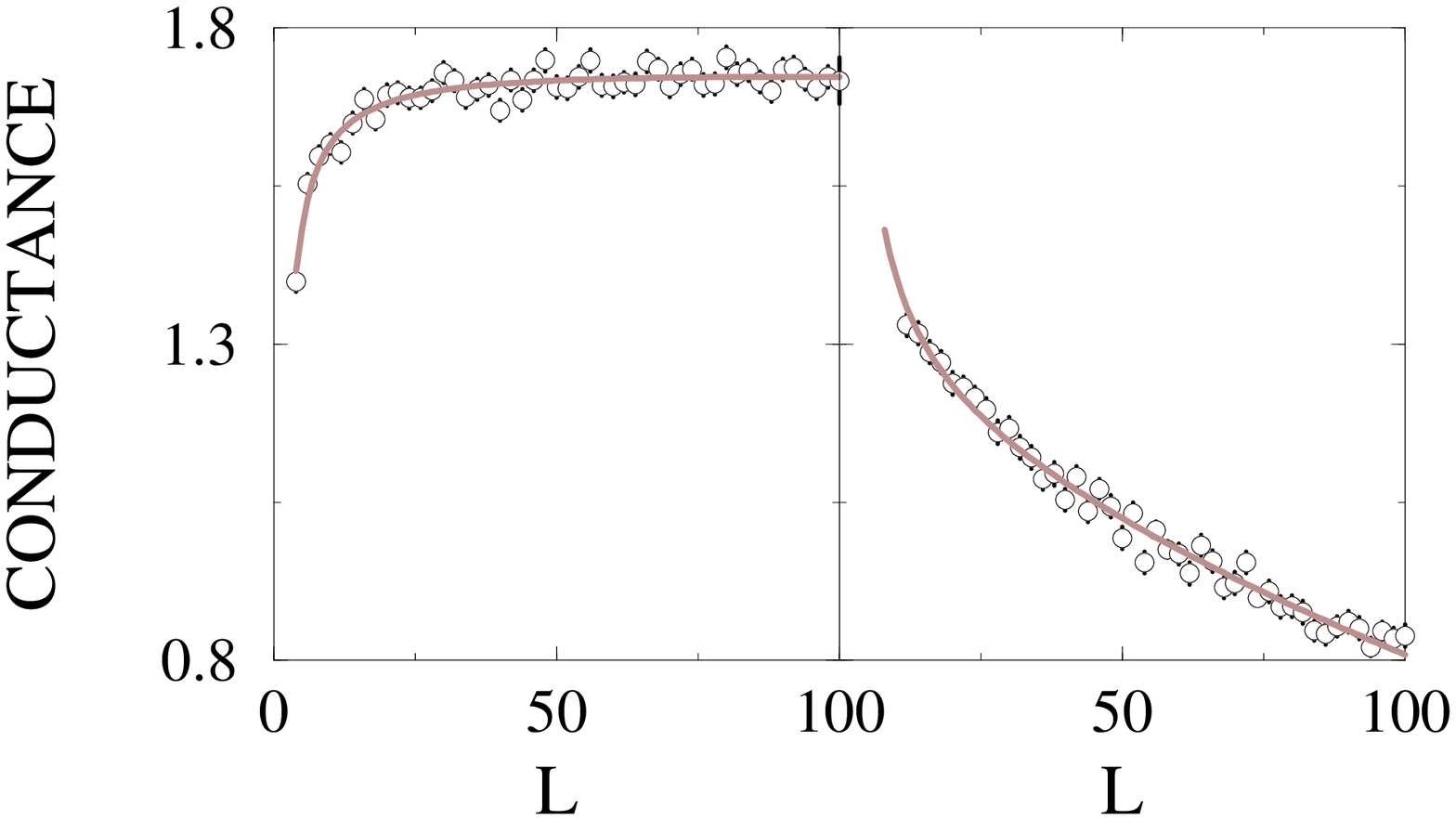,height=6cm,width=8cm}
\caption{Scaling of the average of conductance for the Random
Hopping Sign model at the band center (right part) and at an arbitrary
energy $E=0.1751$ close to it (left part).
The error bars indicate the precision of the mean of the conductance
distribution. The exponential decay corresponds to
a localization length of $\xi = 23317$ at the band center and
$\xi = 461.7$ close to it.}
\label{Figure10.ps}
\end{figure}


\begin{references}
\bibitem{mfp}
In spite of having very large values of the localization length,
the model described by Hamiltonian (\ref{RMF}) gives values of order
1 for the mean free path at energies within the band. This happens for
maximum non-diagonal disorder and $W=0$. Of course, mean free path increases
as disorder diminishes. Most probably, the sharp initial increase of $\xi$
at the band edge is closely related to a similar abrupt increase of
$l$ in the same energy region.
\end{references}
\end{document}